\documentclass[aps, prd, onecolumn, tightenlines, notitlepage, superscriptaddress, nofootinbib, preprintnumbers, floatfix,showkeys,11pt]{revtex4}

\usepackage[normalem]{ulem}
\usepackage{amstext}
\usepackage{amssymb}
\usepackage{amsmath}
\usepackage{graphicx}
\graphicspath{{plots/}}
\usepackage{url}
\usepackage{color}
\usepackage{ulem}
\usepackage[utf8]{inputenc}
\pdfoutput=1
\usepackage{textcomp}

\usepackage{epsfig,amsfonts,mathrsfs,amsmath,amssymb,graphicx,color,slashed,multirow}
\usepackage{amsmath,latexsym,amssymb,graphicx,slashed,color,enumerate,url,cancel,gensymb}
\usepackage{textcomp}

\usepackage[x11names]{xcolor}
\usepackage[colorlinks]{hyperref}

\hypersetup{colorlinks,citecolor= nicered,linkcolor= blue}
\definecolor{nicered}{rgb}{0.7,0.1,0.1}
\definecolor{nicegreen}{rgb}{0.1,0.5,0.1}

\definecolor{blue(ncs)}{rgb}{0.0, 0.53, 0.74}
\AtBeginDocument{\hypersetup{citecolor=blue(ncs),linkcolor=blue,urlcolor=blue(ncs)}}

\usepackage{appendix}

\begin{document}

\title{{\LARGE Signatures of primordial black hole dark matter\\ at DUNE and THEIA}}

\author{Valentina De Romeri}\email{deromeri@ific.uv.es}
\author{Pablo Mart\'inez-Mirav\'e}\email{pamarmi@ific.uv.es}
\author{Mariam T\'ortola}\email{mariam@ific.uv.es}

\affiliation{Departament de F\'isica  Te\'orica,  Universitat  de  Val\`{e}ncia, and Instituto de F\'{i}sica Corpuscular, CSIC-Universitat de Val\`{e}ncia, 46980 Paterna, Spain}

\begin{abstract}
Primordial black holes (PBHs) are a potential dark matter candidate whose masses can span over many orders of magnitude. If they have masses in the $10^{15}-10^{17}$ g range, they can emit sizeable fluxes of MeV neutrinos through evaporation via Hawking radiation. We explore the possibility of detecting light (non-)rotating PBHs with future neutrino experiments. We focus on two next generation facilities:  the Deep Underground Neutrino Experiment (DUNE) and THEIA. We simulate the expected event spectra at both experiments assuming different PBH mass distributions and spins, and we extract the expected 95\% C.L. sensitivities to these scenarios. Our analysis shows that future neutrino experiments like DUNE and THEIA will be able to set competitive constraints on PBH dark matter, thus providing complementary probes in a part of the PBH parameter space currently constrained mainly by photon data.
\end{abstract}

\maketitle

\section{Introduction}
Several astrophysical and cosmological evidences currently indicate that about  $26\%$~\cite{Aghanim:2018eyx} of the content of the Universe is in the form of some cold, non-baryonic matter, called dark matter (DM). The simplest and preferred particle DM models are facing increasingly stronger bounds from a vast array of experimental searches, including colliders and DM direct and indirect detection techniques~\cite{Boveia:2018yeb,Lin:2019uvt,Schumann:2019eaa,PerezdelosHeros:2020qyt,Billard:2021uyg}. In particular, null searches for typical WIMP DM candidates have been fuelling the interest in other possibilities, such as macroscopic objects like black holes (BHs). 
Primordial black holes (PBHs) may have formed via the collapse of large overdensities in the early Universe, originated by an epoch of inflation or by other mechanisms.

PBHs actually constitute one of the earliest proposed and appealing DM candidates~\cite{Zeldovich:1967lct,Hawking:1971ei,Chapline:1975ojl} and, from the point of view of structure formation, they would behave like particle DM on cosmological scales. 
They have recently received renewed attention following the observations of gravitational waves from BH mergers by the LIGO-Virgo collaboration~\cite{Abbott:2016blz,Abbott:2020gyp}, whose origin may be related to PBHs (see for example~\cite{Bird:2016dcv,Clesse:2016vqa,Sasaki:2016jop,Hall:2020daa,Hutsi:2020sol,Franciolini:2021tla}). The theoretically allowed mass range for PBHs spans over many orders of magnitude and thus they can produce a variety of different signals. In turn, a broad number of constraints on the abundance of PBHs have been derived (see e.g. \cite{Sasaki:2018dmp,Carr:2020gox,Carr:2020xqk,Green:2020jor,Villanueva-Domingo:2021spv} for recent reviews), eventually leaving a small mass window (the asteroid-mass one $\sim 10^{17}-10^{21}$ g) open where PBHs could make up all of the DM~\cite{Smyth_2020,Katz:2018zrn,Montero-Camacho:2019jte}.

Black holes are thought to evaporate via Hawking radiation~\cite{Hawking:1974rv,Hawking:1974sw}, emitting particles near their event horizons. While very light non-rotating (maximally rotating) BHs with $M_{\rm BH} \lesssim 5~(7) \times10^{14}$g are expected to have completely evaporated by now~\cite{Page:1976df,Page:1976ki,MacGibbon:2007yq}, those with larger masses would have a lifetime longer than the age of the Universe. 
A large number of constraints have been derived on the abundance of light PBH  ($f_{\rm PBH}=\frac{\Omega_{\rm PBH}}{\Omega_{\rm DM}}$) in the $10^{15}-10^{17}$ g mass range, from the non-observation of their evaporation products: $\gamma$ rays~\cite{1976ApJ...206....8C,Lehoucq:2009ge,Wright:1995bi,Arbey:2019mbc,Ballesteros:2019exr,Laha:2020ivk}, electrons/positrons~\cite{Boudaud_2019,Dasgupta:2019cae,DeRocco:2019fjq,Laha:2019ssq} and neutrinos~\cite{Dasgupta:2019cae}. PBHs in this mass regime are also further constrained by other observations such as the CMB and BBN~\cite{Clark:2016nst,Stocker:2018avm,Acharya:2020jbv,Keith:2020jww}, the EDGES measurements of 21cm absorption \cite{Clark:2018ghm,Hektor:2018qqw,Halder:2021jiv}, or the heating of interstellar medium~\cite{Kim:2020ngi, Laha:2020vhg}. While the most stringent constraints in this mass window currently arise from photons data in the MeV range together with CMB observations and heating of neutral hydrogen \cite{Carr:2009jm,Clark:2018ghm,Kim:2020ngi,Laha:2020ivk,Coogan:2020tuf}, it is intriguing to explore the possibility of probing light PBHs as DM via complementary messengers, like neutrinos.

It has already been shown that a liquid-scintillator detector like that of the future Jiangmen Underground Neutrino Observatory (JUNO) will be able to test non-rotating PBHs abundances as low as $f_{\rm PBH} \sim 10^{-5}$ for masses $M_{\rm PBH} \sim 10^{15}$ g \cite{Wang:2020uvi}, thus improving the current limit set by the non-observation of neutrinos from PBHs evaporation at Super-Kamiokande~\cite{Dasgupta:2019cae}. Moreover, neutrinos from PBH evaporation could be also  observed in DM direct detection experiments through coherent elastic neutrino-nucleus scattering \cite{1867204}.

In this work we investigate the prospects of further exploring the light PBHs parameter space at two future neutrino facilities, the Deep Underground Neutrino Experiment (DUNE)~\cite{Abi:2020evt,Abi:2020wmh} and the proposed THEIA project \cite{Askins:2019oqj}. These experiments will rely on different detector technologies, liquid argon for DUNE and water-based liquid scintillator for THEIA, thus allowing to yield complementary information on the possible detection of MeV neutrinos from PBHs evaporation. Our analysis is also meant to cover a broad range of PBH theoretical scenarios. Besides the simplest case where PBHs follow a monochromatic mass distribution and are non-rotating, we also consider an extended mass distribution and different spin values. Indeed, some cosmological models predict that BHs were already rapidly rotating at the time of their formation (e.g.~\cite{Harada:2017fjm}). A large angular momentum could also be induced in an encounter between two non-rotating black holes \cite{Jaraba:2021ces}. The BH spin parameter is hence a fundamental property, which plays a relevant role in its formation and can affect its characteristic detection signatures~\cite{Arbey:2019vqx,Dasgupta:2019cae,Kuhnel:2019zbc,Ray:2021mxu}. We will show that both DUNE and THEIA have the potential to improve the current bound set by Super-Kamiokande \cite{Dasgupta:2019cae} and allow to probe PBH abundances close to current constraints from $\gamma$ rays and other observations. Remarkably, even if their primary physics goal is the study of neutrino properties and flavour oscillations, these next-generation neutrino experiments will allow to provide relevant information on the nature of DM.\\

The rest of the paper is organised as follows. In Section~\ref{sec:PBHevap}, we introduce the formalism to compute neutrino fluxes from PBH evaporation.
The details on the simulation of the event spectra in DUNE and THEIA are provided in Section~\ref{sec:events}.
We derive the expected sensitivities to the PBH scenarios in Section~\ref{sec:results} and finally draw our conclusions in Section~\ref{sec:conclusions}.

\section{Neutrinos from primordial black holes evaporation}
\label{sec:PBHevap}

A rotating BH emits elementary particles following a blackbody-like distribution with a temperature~\cite{Page:1976df,Page:1976ki,MacGibbon:1990zk,MacGibbon:1991tj,MacGibbon:2007yq}
\begin{equation}
    T = \frac{1}{4\pi M_{\rm BH} G_N} \frac{\sqrt{1-a_*^2}}{1+\sqrt{1-a_*^2}},
\end{equation}
where $G_N$ is the gravitational constant, $M_{\rm BH}$ is the BH mass and $a_*$ is the reduced spin parameter, $a_*=\frac{J_{\rm BH}}{G_N M_{\rm BH}^2}$, with $J_{\rm BH}$ the BH angular momentum. The BH emits significant numbers of particles with energies close to its temperature $T$, while the emission becomes exponentially suppressed for particle energies larger than $T$.

The number of $i$ particles with spin $s$ emitted from an evaporating BH is~\cite{Hawking:1971ei,Page:1976df,Page:1976ki,Page:1977um,MacGibbon:1990zk,MacGibbon:1991tj,MacGibbon:2007yq}
\begin{equation}
     \frac{d^2 N_i}{dE dt} = \sum_{\text{dof}} \frac{1}{2\pi}\frac{\Gamma_i (E,M_{\rm BH},a_*,m_0,s)}{e^{E'/T} - (-1)^{2s}},
\end{equation}
where the sum is over the multiplicity of the particles (color, helicity, angular momentum and its projection $m$), $E'$ is the effective energy of the emitted particles, $E' = E - m\Omega$, including the BH angular velocity $\Omega = \frac{a_*}{2G_N M_{BH}}\left(\frac{1}{(1 + \sqrt{1 - a_*^2})}\right)$. 
The function $\Gamma_i$ is the so-called greybody factor, which encodes the probability of an elementary particle to escape the BH gravitational well. It depends on the BH mass, the BH spin and the rest mass and spin of the emitted particle, $m_0$\footnote{In the case of a Schwarzschild BH, all particles with equal angular momenta $l$ give the same contribution to $\Gamma_i$.}.

In full generality, we expect two contributions to the neutrino spectra generated by Hawking radiation of the distribution of PBHs. The first one is the \textit{primary} contribution  consisting of neutrinos directly emitted in the evaporation. The second one is the \textit{secondary} contribution, which stems from the hadronization and further decay of the primary particles.
The total differential number of particles emitted  from an evaporating PBH is given by the sum of the two contributions. We also expect both PBHs in the galactic halo and outside to contribute to the differential neutrino flux from PBH evaporation: 
\begin{equation}
 \frac{d\phi^\nu_{\rm tot}} {dE}= \frac{d\phi^\nu_{\rm gal}} {dE} +
 \frac{d\phi^\nu_{\rm exg}} {dE}.
 \label{eq:totaldiffflux}
 \end{equation}

Assuming a monochromatic PBH mass distribution, the galactic contribution reads
\begin{equation}
 \frac{d\phi^\nu_{\rm gal}} {dE}=  \frac{f_{\rm PBH}}{M_{\rm{PBH}}}   \frac{d^2N}{dE dt} \int \frac{1}{4 \pi} d \Omega \int \rho_{\rm MW}\,[r(\ell,\psi)]\,d\ell \,,
 \label{eq:dFdEgal}
 \end{equation}
where $\rho_{\rm MW}\,[r(\ell,\psi)]$ denotes the DM profile of the Milky Way (MW) halo and $f_{\rm PBH}$ is the fraction of DM composed of PBHs. Moreover, $r(\ell,\psi) = \sqrt{d_{\odot}^2 + \ell^2 - 2\ell d_\odot \rm cos(\psi)}$ is the galactocentric distance, with $\ell$ the line of sight (l.o.s.), $d_\odot$ the galactocentric distance of the Sun, $\psi$ the angle of view which defines the l.o.s. and $d \Omega$ the differential solid angle considered. For the sake of illustration and throughout our analysis, we assume a Navarro-Frenk-White (NFW) density profile~\cite{Navarro:1995iw} for the MW halo (with scale radius $r_s = 20$ kpc and normalisation $\rho_0 = 0.4~ \rm GeV/cm^3$). We have checked that the final sensitivities do not depend substantially on this choice. For instance, they vary by a factor of 2 when considering a cored DM density profile, namely an isothermal with  $r_s = 1.5$ kpc.

The subdominant extragalactic contribution is obtained under the assumption that the DM is distributed isotropically at sufficiently large scales. The corresponding differential neutrino flux over the full sky is estimated as the redshifted sum over all epochs emission~\cite{Carr:2009jm,Arbey:2019vqx,Dasgupta:2019cae,Wang:2020uvi}:
\begin{equation}
\frac{d\phi^\nu_{\rm exg}} {dE}= 
\frac{f_{\rm{PBH}}\,\bar{\rho}_{\rm DM}}{M_{\rm{PBH}}}  \int_{t_{\rm min}}^{t_{\rm max}} dt [1+z(t)]\, \frac{d^2N}{dE_0 dt}\Bigr\rvert_{E_0 = [1+z(t)]E}\,,
\label{eq:dFdEexg}
\end{equation}
where $\bar{\rho}_{\rm DM} = 2.35\times 10^{-30}$ g cm$^{-3}$ is the average DM density today~\cite{Aghanim:2018eyx} and the integral limits are: $t_{\rm min} = 1 s$\footnote{Note that changing this lower limit has essentially no impact on the results~\cite{Dasgupta:2019cae,Wang:2020uvi}.} (neutrino decoupling time) and  $t_{\rm max}$ is the minimum value between the PBH lifetime ($\propto M_{\rm{PBH}}^3$) and the age of the Universe. The neutrino energy at the source $E_0$ is related to the neutrino energy in the observer's frame via the redshift parameter $z(t)$. We fix all cosmological parameters according to the last Planck results~\cite{Aghanim:2018eyx}. 

\begin{figure}[!tb]
\centering
\includegraphics[width=0.45\textwidth]{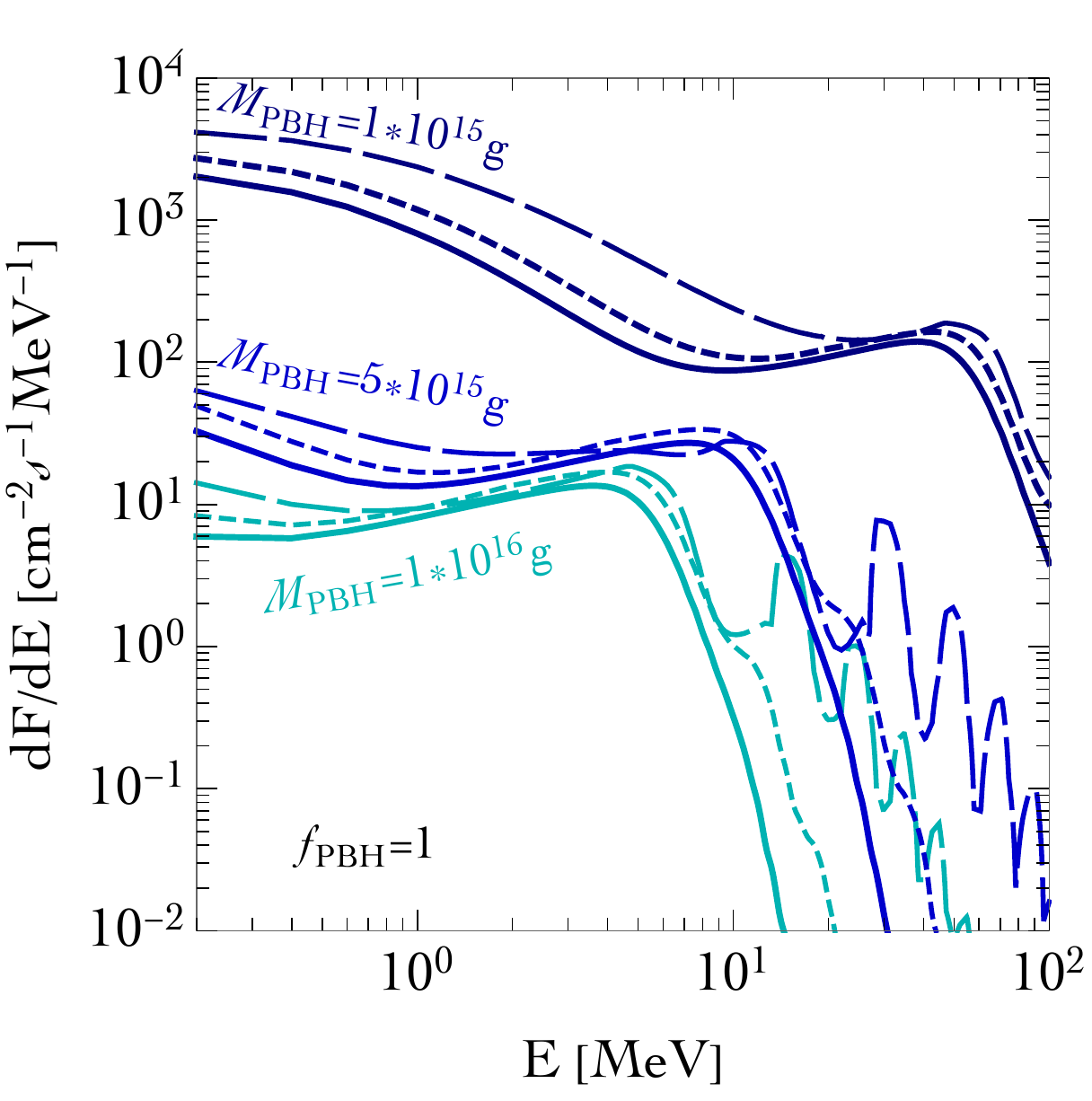}
\caption{Total differential $\nu_e$ flux from PBHs, for three different PBH masses ($M_{\rm{PBH}} = 1\times 10^{15}$ g (dark blue), $5\times 10^{15}$ g (blue) and $1\times 10^{16}$ g (cyan)) and spins ($a_*=0$ (solid), $a_*=0.5$ (dashed) and $a_*=0.9$ (large dashed)). A monochromatic mass distribution of PBHs  and $f_{\rm PBH} =1$ have been assumed. }
\label{fig:totaldFdE}
\end{figure}

 Fig.~\ref{fig:totaldFdE} shows the total (galactic + extragalactic, primary + secondary) differential $\nu_e$ flux from PBH evaporation expected on Earth, for three different masses $M_{\rm{PBH}} = 1\times 10^{15}$ g (dark blue), $5\times 10^{15}$ g (blue) and $1\times 10^{16}$ g (cyan). We  assume a monochromatic PBH mass distribution and different PBH spins (also following a monochromatic distribution): $a_*=0$ (solid), $a_*=0.5$ (dashed) and $a_*=0.9$ (large dashed). For simplicity and illustration purposes, we consider the limit case in which  PBHs constitute the totality of the DM ($f_{\rm PBH} =1$). 
Heavier PBHs lead to reduced fluxes, shifted towards smaller neutrino energies. If the PBH has a large spin parameter, the angular momentum of the particles emitted plays an important role and leads to an emitted flux with evident peaks, as shown in Fig.~\ref{fig:totaldFdE}. The dominant (smooth) contribution to the flux stems from particles emitted with no additional angular momentum apart from their intrinsic spin. The characteristic peaks at higher energies correspond to the radiation of particles with higher angular momentum, leading to a substantial enhancement of the flux if the latter is aligned with the PBH one (see for instance \cite{Page:1976df,Page:1976ki} for a detailed discussion).

While a PBH monochromatic mass distribution has the advantage of simplicity, a large variety of formation mechanisms actually predicts extended mass distributions for PBHs. In such scenarios, constraints are expected to be modified compared to monochromatic distributions~\cite{Carr:2017jsz,Kuhnel:2017pwq,Bellomo:2017zsr,Arbey:2019vqx,Dasgupta:2019cae}. In full generality, one should compute the pattern of density perturbations induced by a specific formation scenario (for instance from inflation) and use this quantity to determine 
the mass distribution of PBHs generated by the critical collapse of these inhomogeneities. From a phenomenological point of view, instead of complying with a specific formation mechanism or scenario, one can adopt a simple parametric function for the PBH mass distribution.
The most common approximation to the true mass distribution generated from a peak in the power spectrum of primordial fluctuations is the log-normal distribution~\cite{Carr:2017jsz} for the comoving number density:
\begin{equation}
\frac{dN_{\rm{PBH}}}{dM_{\rm{PBH}}}= \frac{1}{\sqrt{2\pi}\sigma M_{\rm{PBH}}}\, \exp\left[- \dfrac{ {\rm ln} \left(M_{\rm PBH}/M_c\right)^2}{2 \sigma^2} \right]\,,
\label{eq: log-normal mass distribution}
\end{equation} 
where $M_c$ is the critical mass (referring to the position of the peak) and $\sigma$ is the width of the distribution.  The distribution is normalised to 1 when the lower and upper integration  limits tend to 0 and infinity, respectively\footnote{Although this condition can not be exactly met numerically, we sample over a large population of PBHs with masses spanning over more than two orders of magnitude, in order to ensure the validity of the result.}. The exact values of $\sigma$ and $M_c$ will depend on the specific PBH formation scenario. In our analysis, we let $M_c$ vary in the range of $M_{\rm{PBH}}$ that can be explored with neutrino experiments, and allow for three different values of $\sigma$, $\sigma = 0.1, 0.5$ and 1 to account for both narrow and broad mass functions.

\begin{figure}[!tb]
    \centering
    \includegraphics[width = 0.45\textwidth]{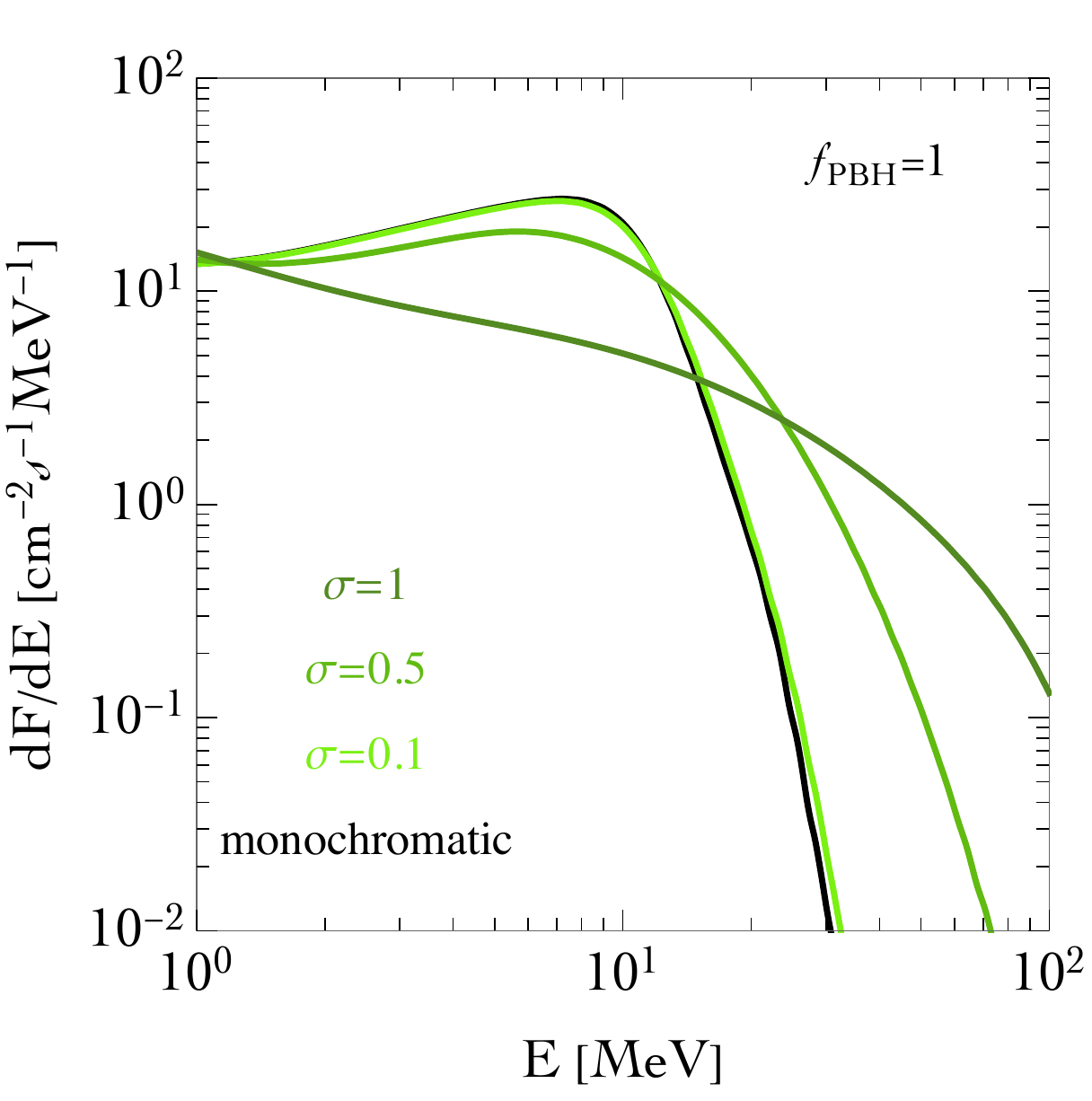}
    \caption{Comparison between the total neutrino flux expected from PBH evaporation, for a monochromatic mass distribution with $M_{\rm PBH} = 5 \times 10^{15}$ g (black) and a log-normal distribution with critical mass $M_c = 5 \times 10 ^{15}$ g and $\sigma = \lbrace 0.1, 0.5, 1 \rbrace$ (light, medium and dark green). The fraction of DM consisting of PBH is fixed to $f_{\rm PBH} = 1$ and  $a_*=0$.}
    \label{fig:lognorm}
\end{figure}

In this case, the fluxes of galactic and extragalactic neutrinos are given by
\begin{equation}
 \frac{d\phi^\nu_{\rm gal}} {dE}=  \frac{f_{\rm PBH}}{\overline{M}_{\rm{PBH}}} \int \frac{1}{4 \pi} d \Omega \int \rho_{\rm MW}\,[r(\ell,\psi)]\,d\ell \int_{M_{\rm min}}^{M_{\rm max}} \frac{dN_{\rm{PBH}}}{dM_{\rm{PBH}}}\, \frac{d^2N}{dE dt} d M_{{\rm PBH}}\,,
 \label{eq:dFdEgal-lognorm}
 \end{equation}
 and
\begin{equation}
\frac{d\phi^\nu_{\rm exg}} {dE}= 
\frac{f_{\rm{PBH}}\,\bar{\rho}_{\rm DM}}{\overline{M}_{\rm{PBH}}}  \int_{t_{\rm min}}^{t_{\rm max}} dt [1+z(t)]\, \int_{M_{\rm min}}^{M_{\rm max}} \frac{dN_{\rm{PBH}}}{dM_{\rm{PBH}}} \frac{d^2N}{dE_0 dt}\Bigr\rvert_{E_0 = [1+z(t)]E}\,dM_{{\rm PBH}}\,,
\label{eq:dFdEexg-lognorm}
\end{equation}
where $\overline{M}_{\rm PBH}$ denotes the mean mass of the distribution in Eq.~\ref{eq: log-normal mass distribution} . 
The integration intervals are chosen in order to ensure that the relevant range of masses is covered.  The impact of considering such a distribution can be seen in Figure \ref{fig:lognorm}, where the expected differential neutrino fluxes from evaporating PBHs with a log-normal mass distribution are shown, assuming different values of the standard deviation, and compared to the case of a monochromatic distribution. As one can see, the larger the distribution width, the more the typical ''peak" from the galactic primary contribution is smoothed down and, as a result, the flux is still sizeable at larger energies.
 
We rely on the public tool \texttt{BlackHawk} \cite{Arbey:2019mbc} to compute both the instantaneous and the full time-dependent emission of neutrinos from a black hole of a given mass and spin, including the primary and secondary contributions. Hadronization is done using \texttt{PYTHIA}~\cite{Sjostrand:2014zea} at the present epoch (galactic contribution) or in the early Universe (extragalactic contribution). It is important to point out that the secondary component of the neutrino flux at low energies is not computed directly with \texttt{PYTHIA} but it is extrapolated from the calculations at higher energies. Although this extrapolation is a source of uncertainty, there is not a more precise computation of the neutrino flux publicly available at present \cite{Arbey:2021mbl}. In this work, due to the PBH mass range that we will be exploring and the window of neutrino energies accessible, the primary component will be the dominant contribution to the flux and only the sensitivity to $M_{\rm PBH} \sim 10^{15}$ g might be partially affected by the extrapolation.

\begin{figure}[!tb]
\centering
\includegraphics[width=0.45\textwidth]{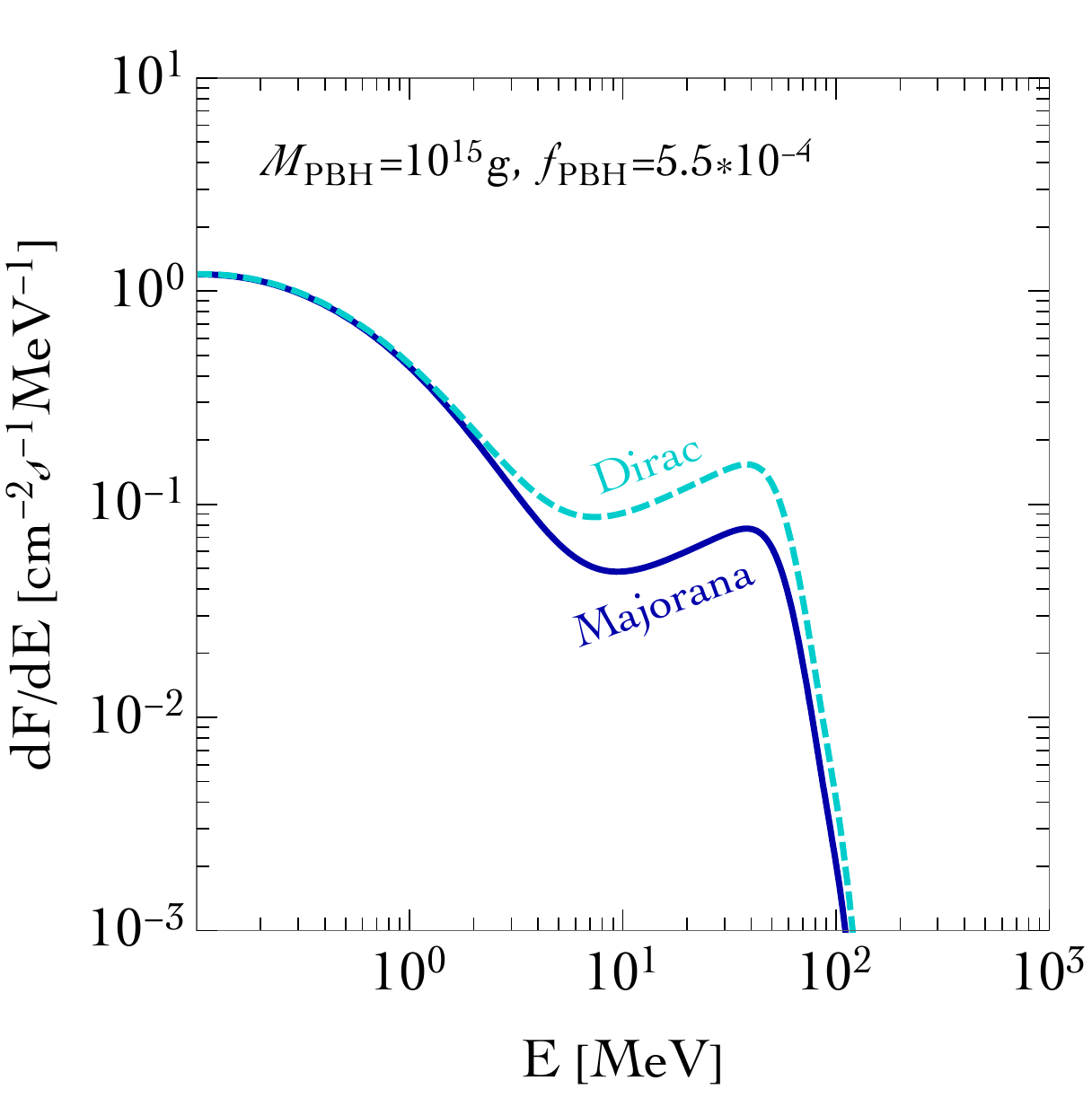}
\caption{Total differential $\nu_e$ flux from PBHs (primary + secondary contribution), where $M_{\rm PBH}=10^{15}$ g, $f_{\rm PBH}=5.5 \times 10^{-4}$ (upper limit from Super-Kamiokande~\cite{Dasgupta:2019cae}), $a_*=0$ and  a monochromatic mass distribution of PBHs are assumed. The blue, solid line is for Majorana neutrinos, while the cyan, dashed curve refers to Dirac neutrinos.}
\label{fig:vDiracvsMaj}
\end{figure}

The PBH evaporation rate depends on the total number of degrees of freedom of the emitted particles, as well as on their masses, and could be sensitive to extensions of the Standard Model \cite{Baker:2021btk}.
For simplicity, in the code and throughout the analysis, neutrinos are considered to be massless (non-zero masses would make a very small difference, as they would be relevant mostly when $T \sim m_\nu$~\cite{Lunardini:2019zob}) and with a Majorana nature (6 degrees of freedom).  A  slightly larger differential flux would be expected in the case of Dirac neutrinos (12 degrees of freedom)~\cite{Lunardini:2019zob}.
We show in Fig.~\ref{fig:vDiracvsMaj} the expected total differential $\nu_e$ flux from PBHs  (primary + secondary contribution) for a monochromatic mass distribution with $M_{\rm PBH}=10^{15}$ g and fixing $f_{\rm PBH}=5.5 \times 10^{-4}$ (current upper limit from Super-Kamiokande~\cite{Dasgupta:2019cae}). The blue, solid line assumes Majorana neutrinos, while the cyan, dashed curve refers to Dirac neutrinos. The addition of new degrees of freedom in the Dirac neutrino case increases the expected flux up to a factor 2 close to the peak~\cite{Lunardini:2019zob}, in the PBH mass range of interest here. Let us notice however, that pragmatically a neutrino detector would not be able to distinguish the nature of neutrinos, as only left-handed helicity eigenstates would be seen. \\

Let us also mention that we do not include the effect of neutrino mixing in our analysis. The primary component of the flux, arising directly from evaporation, is not affected since neutrinos are emitted as mass eigenstates and the impact of their masses has been shown to be negligible in this phenomenon~\cite{Lunardini:2019zob}. Hence, flavour conversions would only affect the secondary components of both the galactic and the extragalactic contributions to the total flux, since neutrinos are produced in the decay of heavier particles as flavour eigenstates. Then, including neutrino mixing, the secondary flux expected on Earth reads
\begin{equation}
    \label{eqn:osc}
    \left. \frac{d\phi^{\nu_e}_{\rm sec}}{dE} \right\vert_{\rm detector} = \sum_{\alpha } P (\nu_\alpha \rightarrow \nu_e)   \frac{d\phi^{\nu_\alpha}}{dE} \approx \sum_{\alpha, j} |U_{\alpha j}|^2 |U_{ej}|^2  \frac{d\phi^{\nu_\alpha}}{dE},
\end{equation}
where the Greek and Latin indices refer to flavour and mass eigenstates, respectively. We have checked numerically that considering neutrino mixing \footnote{The values of the oscillation parameters were taken from \cite{deSalas:2020pgw}.} would reduce the total differential flux by less than $\sim 2$\% at the lowest energies we are interested in and for the range of PBH masses we are exploring. Therefore, the inclusion of neutrino mixing would have a negligible impact on the results presented in this work. Nonetheless, note that flavour mixing can induce $\mathcal{O}(10\%)$ changes in the expected flux at energies much below the dominant peak generated by the primary contribution with galactic origin. \\

\section{Event spectra at future facilities}
\label{sec:events}

As already shown in Fig.~\ref{fig:totaldFdE}, the expected neutrino flux from PBH evaporation can be sizeable in magnitude (if $f_{\rm PBH}=1$), and centered around MeV neutrino energies, provided that $M_{\rm PBH}\sim 10^{15} -10^{16}$ g. Heavier, asteroid-mass PBHs, while more interesting as DM candidates as currently unconstrained, would produce a weaker flux of neutrinos with sub-MeV energies. Their detection is therefore unfeasible also given the huge background of solar, reactor and geoneutrinos at the same energies. Moreover, if PBHs constitute only a fraction of the total DM content of the Universe, such a flux will be reduced by a total factor proportional to $f_{\rm PBH}$. 
Going back to the MeV range, small fluxes of MeV neutrinos from PBH evaporation may be detected at future neutrino experiments.
Large cross sections, large exposures and low or well-controlled backgrounds are essential ingredients. In this work, we focus on DUNE, a liquid argon time-projection chamber (LArTPC) and THEIA, a proposed water-based liquid scintillator neutrino detector. Other present and near-future experiments such as the loading of gadolinium in Super-Kamiokande~\cite{Beacom:2003nk, Kibayashi:2009ih} and Hyper-Kamiokande~\cite{Abe:2018uyc} or reactor experiments like JUNO~\cite{Wang:2020uvi,Abusleme:2021zrw} will also be sensitive to the predicted flux of neutrinos with PBH origin.  

\subsection{DUNE}

The far detector of the future DUNE experiment will be a very large, modular LArTPC with a total fiducial mass of at least 40 kton~\cite{Abi:2020evt}. 
In principle, several independent channels are accessible with a LArTPC: elastic scattering on electrons, $\nu_e$ and $\bar{\nu}_e$ charged-current (CC) absorption in $^{40}$Ar and neutral-current interactions on $^{40}$Ar.
The largest cross section is that of $\nu_e$-CC interactions $\nu_e + ^{40}$Ar $\rightarrow e^- + ^{40}K^*$.
Hence the DUNE far detector, contrarily to the case of water Cherenkov or liquid scintillator detectors, will be mainly sensitive to the $\nu_e$ component of the PBH neutrino flux.  The  predicted $\nu_e$  event  rate  from  PBH evaporation is then calculated  by folding the  expected  neutrino  event rate  with the $\nu_e + ^{40}$Ar $\rightarrow e^- + ^{40}K^*$ cross section and the DUNE detector parameters using \texttt{SNOwGLoBES}~\cite{snowglobes}, a fast event-rate computation  tool which employes the front-end rate engine part of \texttt{GLoBES}~\cite{Huber:2004ka,Huber:2007ji}. Sticking to the \texttt{SNOwGLoBES} configuration for liquid argon (LAr) detectors,  we assume a detection threshold of 5 MeV and the following energy resolution for the smearing matrices~\cite{snowglobes,Amoruso:2003sw} 
\begin{equation}
    \left(\frac{\sigma}{E}\right)^{2} = \left(\frac{0.11}{\sqrt{E~({\rm MeV})}}\right)^{2}~+~(0.02)^{2} \,. 
\end{equation}
The $\nu_e$-CC absorption in $^{40}$Ar produces an electron and a cascade of photons from the deexcitation of potassium, which may also help to tag the $\nu_e$ channel. Let us notice that current estimations for the $\nu_e$-CC absorption cross section are affected by uncertainties at the 10-20\% level~\cite{snowglobes}. For our calculations, we stick to the estimations from~\cite{GilBotella:2003sz}. We have checked numerically that using an alternative cross section for CC interactions on argon, namely from the event generator MARLEY~\cite{Gardiner2021,MARLEYv1.2.0,marleyPRC}, would lead to a depletion of the number of events at $E \gtrsim 50$ MeV. In particular we find a factor $\sim 1.5$ less events at $E=100$ MeV, for $M_{\rm PBH}=10^{15}$ g, $f_{\rm PBH}=1 \times 10^{-4}$, $a_*=0$ and assuming a monochromatic mass distribution of PBHs. Moreover, for simplicity and giving the lack of available information, we further assume $100\%$ efficiency in the detection. 

Several backgrounds make the detection of neutrinos from PBH evaporation challenging. At DUNE, solar neutrinos (at low energies, $\lesssim 16$ MeV) and atmospheric neutrinos (at high energies, $\gtrsim 30$ MeV) constitute the dominant and irreducible backgrounds. We will therefore restrict our analysis to an energy window between 16 MeV and 100 MeV so that the large solar neutrinos background does not have to be included in our simulation.
On the other hand, the expected atmospheric neutrino events are computed using \texttt{SNOwGLoBES} as well. We consider the prediction for the atmospheric neutrino flux calculated with a FLUKA simulation for the Gran Sasso laboratory~\cite{Battistoni:2005pd}. Since the latitude of this laboratory is similar to that of the Sanford Underground Research Facility (SURF), where the DUNE far detector will be placed, one does not need to worry about the latitude dependence of the flux. The flux predicted with FLUKA in the range of energies between 100 MeV and 300 MeV is then rescaled to the HKKM atmospheric neutrino flux \cite{Honda:2015fha} in the same energy range at the right location, the Homestake mine. The uncertainty of this low-energy atmospheric neutrino flux prediction is about 35$\%$ \cite{Battistoni:2005pd, Guo:2018sno,Sawatzki:2020mpb}. Finally, the diffuse supernova neutrino background (DSNB) constitutes a background too. While not yet observed, DUNE far detector would be sensitive to the $\nu_e$ component of this diffuse relic supernova neutrino flux \cite{Moller:2018kpn,Abi:2020evt}. We calculate the DSNB flux following~\cite{deGouvea:2020eqq} and assuming an estimated uncertainty of 35$\%$ due to the uncertainty in the star formation rate. As for the atmospheric background, we simulate the interaction of the DSNB flux with LAr using \texttt{SNOwGLoBES}.

\subsection{THEIA}

A water-based liquid scintillator (WbLS) can be obtained by adding ultrapure water to an organic scintillator. The goal is to obtain a transparent medium so that, with the suitable instrumentation, it is possible to detect Cherenkov light and scintillation photons simultaneously \cite{Bignell:2015oqa, Caravaca:2020lfs}. THEIA is a proposal for a future very large detector using such technology \cite{Askins:2019oqj}. There are two possible setups currently under consideration: a 25 kton tank placed at SURF, that could act as a fourth tank for DUNE; or a larger tank of 100 kton whose location is still undetermined. For the detection of supernova neutrinos it has been shown \cite{Askins:2019oqj} that the dominant reaction channel in THEIA will be inverse beta decay (IBD): $\bar{\nu}_e + p \rightarrow e^+ + n$. Both final states, $e^+$ and $n$, will be visible in the WbLS. 
In our analysis we focus on this main detection channel, implementing the corresponding cross section from \cite{Strumia:2003zx}. We choose a conservative energy resolution of $7\% / \sqrt{E \text{ (MeV)}}$ and we will assume the fiducial volumes to be 20 kton ($1.55 \times 10^{33}$ targets) and 80 kton ($6.2 \times 10^{33}$ targets) respectively, for each of the two possible configurations. 

Different backgrounds will be relevant in THEIA too. In addition to the CC atmospheric neutrino background expected at high energies and the DSNB, as for DUNE, other sources of backgrounds are expected. At low energies, reactor antineutrinos constitute an irreducible background. We exclude them from our analysis by focusing on an energy window starting at 10 MeV, and which we will extend up to 100 MeV. Neutral current (NC) interactions of high-energy atmospheric neutrinos in the detector, in which a neutron is released, can mimic IBD events. This background can be enormously reduced with the appropriate cuts and benefiting from the WbLS capability to separate Cherenkov and scintillation light. In our analysis, we consider the expected reduced background obtained in a DSNB-dedicated analysis for THEIA \cite{Sawatzki:2020mpb}, where the impact of applying the appropriate cuts results in an 80$\%$ efficiency. \\ 

\begin{figure}[!tb]
\centering
\includegraphics[width=0.7\textwidth]{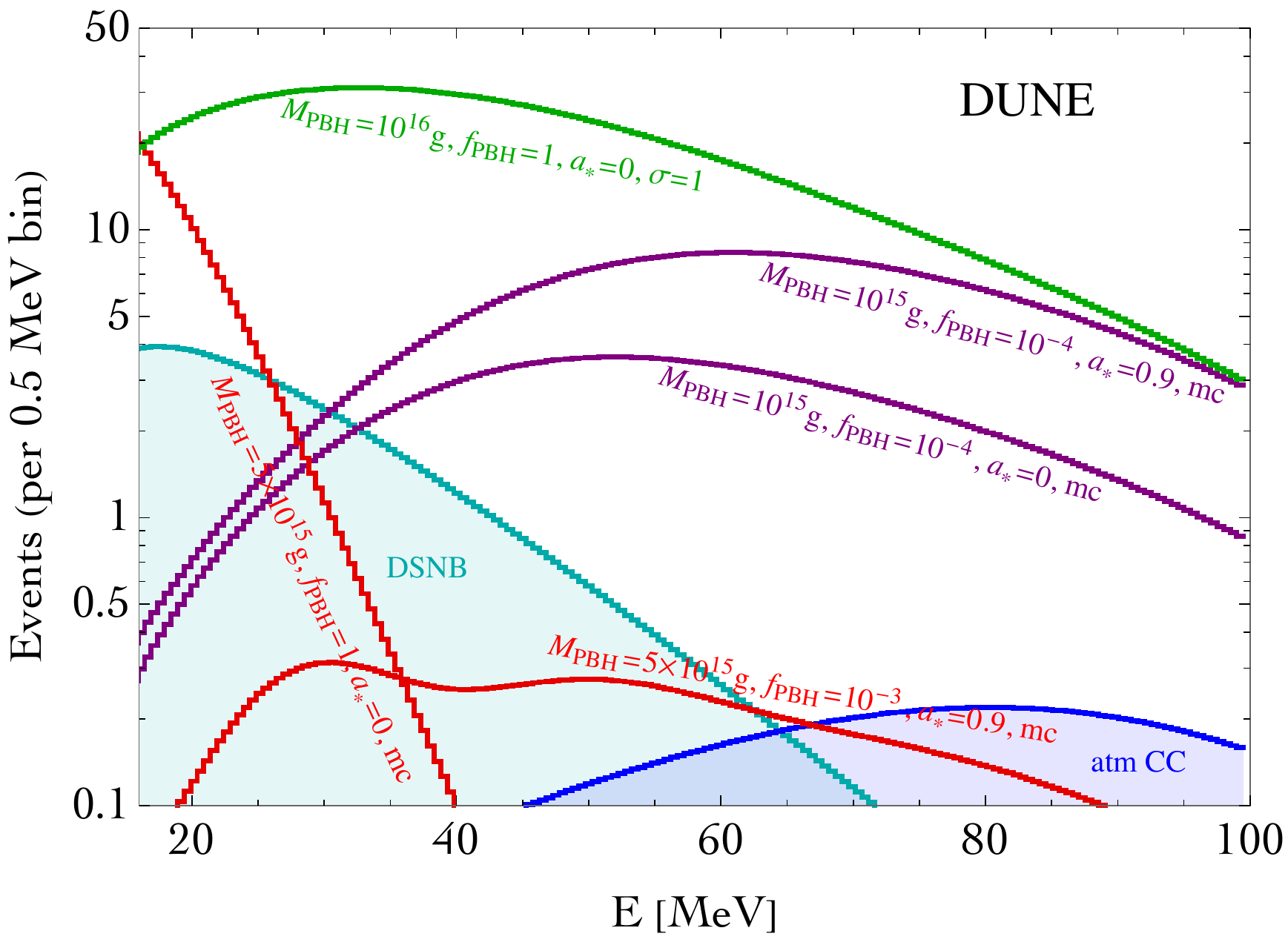}
\includegraphics[width=0.7\textwidth]{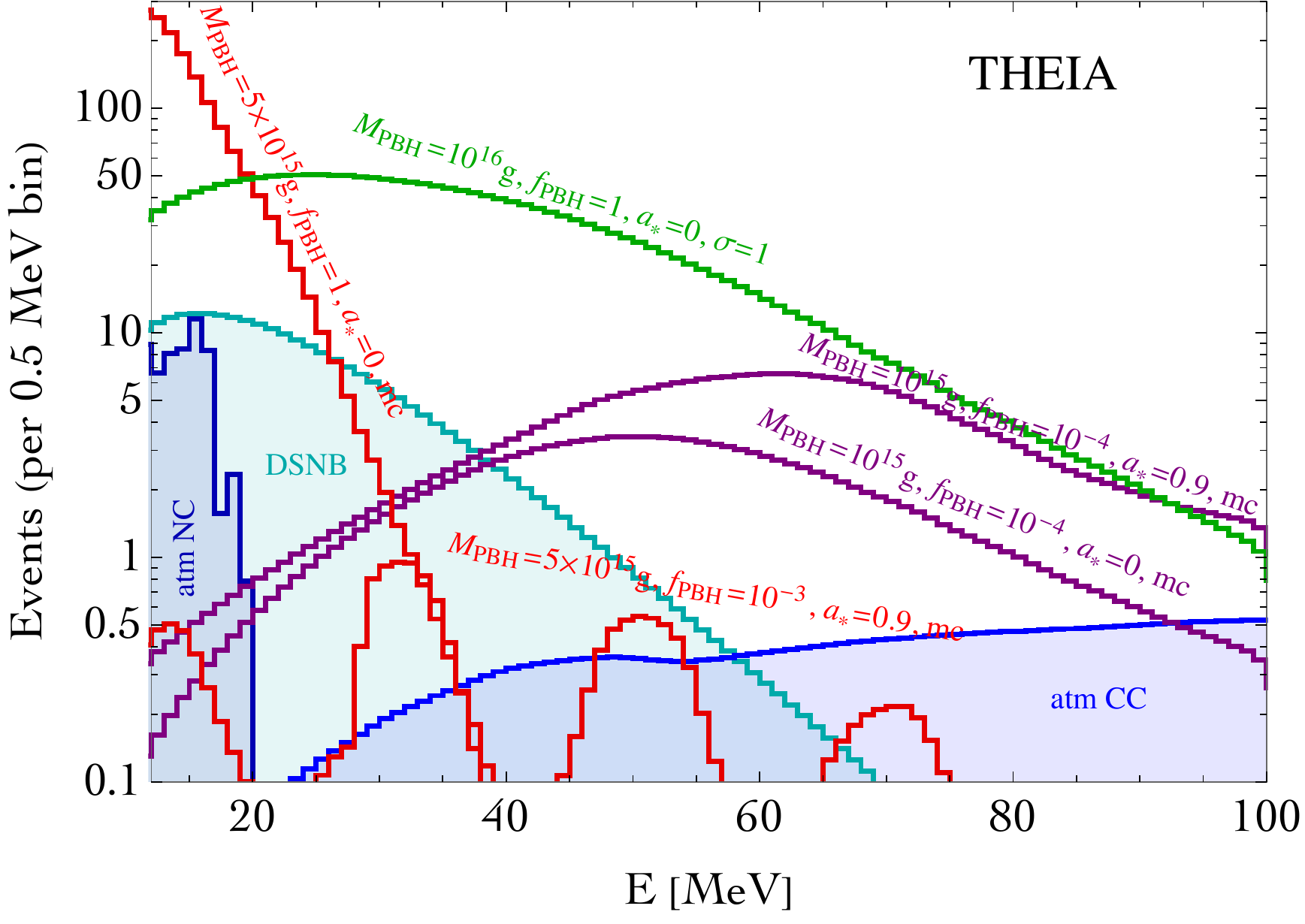}
\caption{Number of events as a function of the neutrino energy at DUNE (upper panel) and THEIA (lower panel). For DUNE we assume the main detection channel
$\nu_e + ^{40}$Ar $\rightarrow e^- + ^{40}K^*$, 40 kton of fiducial volume for the far detector, 10 years of operation and $0.5$ MeV energy bins. For THEIA, we focus on the IBD detection channel and we take 80 kton fiducial volume for detector, 10 years of operation and $1$ MeV energy bins. Different PBH masses and spins, as well as DM fractions are considered as described in the legends. Expected background events from DSNB, CC and NC atmospheric neutrino interactions are also shown as for comparison.}
\label{fig:events_spectra}
\end{figure}

We show in Fig.~\ref{fig:events_spectra} the predicted number of events as a function of the neutrino energy in DUNE (upper panel) and THEIA (lower panel). In the case of DUNE, and as previously discussed, we consider the main detection channel of $\nu_e$-CC interactions $\nu_e + ^{40}$Ar $\rightarrow e^- + ^{40}K^*$. We assume a 40 kton fiducial volume for the far detector, 10 years of operation and $0.5$ MeV energy bins. The cyan area indicates the expected background events from DSNB, while the blue one those from the atmospheric neutrino charged-current background. For THEIA, instead, we assume IBD as the main detection channel, 80 kton fiducial volume for the detector, 10 years of operation and $1$ MeV energy bins. The additional background arising from NC interactions of high-energy atmospheric neutrinos is shown in dark blue.
In both panels, the coloured curves depict the predicted spectra assuming different PBH masses, spins, and dark matter fractions chosen as examples (see legends in the plots). All of them assume a monochromatic (mc) mass distribution except for the green curve which assumes a log-normal mass distribution with $\sigma=1$.

Let us also recall that the number of events scales linearly with the DM fraction $f_{\rm PBH}$ and shifts towards lower neutrino energies as the PBH mass increases. 
The differences in the detection channel, and hence, in the cross section, as well as the different energy resolutions, lead to non-identical event spectra. For instance, THEIA would be capable of resolving the characteristic peaks of a PBH population with large spin ($a_* = 0.9$) whereas, in DUNE, this typical signature would be smeared out.\\

\section{Sensitivities at DUNE and THEIA}
\label{sec:results}

In this section, we extract the projected sensitivities for the detection of neutrinos from PBH evaporation in DUNE and THEIA.
For DUNE, we use the following least-square function:
\begin{equation}
  \label{eq:chiSqDUNE}
  \chi_{\rm DUNE}^2=\sum_{i}\left(\frac{N^\text{\rm PBH}_i
    +(1+\alpha)N^\text{\rm atmCC}_i
    +(1+\beta)N^\text{\rm DSNB}_i-N^\text{\rm atmCC}_i-N^\text{\rm DSNB}_i}{\sqrt{N^\text{\rm PBH}_i+N^\text{\rm atmCC}_i+N^\text{\rm DSNB}_i}} \right)^2
  + \left(\frac{\alpha}{\sigma_\alpha}\right)^2
  + \left(\frac{\beta}{\sigma_\beta}\right)^2\ ,
\end{equation}
where $N^\text{\rm PBH}_i$ refers to the predicted number of neutrino events from PBH evaporation in the energy bin $i$, $N^\text{\rm DSNB}_i$ are the backgrounds events from the DSNB and $N^\text{\rm atmCC}$ the events from the atmospheric neutrino charged-current background. The nuisance parameters $\alpha$ and $\beta$
account for uncertainties on the
backgrounds. We fix their standard deviations to $\sigma_\alpha=0.35$ and
$\sigma_\beta=0.35$.
In the case of THEIA, there is an additional background due to atmospheric neutrinos interacting via neutral currents, which can mimic an IBD event. We modify Eq. \ref{eq:chiSqDUNE} accordingly to account for this background, including an additional pull parameter related to the normalisation of the atmospheric neutral current events, whose systematic uncertainty $\sigma_\gamma$ we fix to 30\%~\cite{Collaboration:2011jza,PhysRevD.103.053002}:

\begin{multline}
  \label{eq:chiSqTHEIA}
  \chi_{\rm THEIA}^2 = \\  \sum_{i}\left(\frac{N^\text{\rm PBH}_i
    +(1+\alpha)N^\text{\rm atmCC}_i
    +(1+\beta)N^\text{\rm DSNB}_i 
    +(1+\gamma)N^\text{\rm atmNC}_i 
    -N^\text{\rm atmCC}_i
    -N^\text{\rm DSNB}_i
    -N^\text{\rm atmNC}_i}{\sqrt{N^\text{\rm PBH}_i+N^\text{\rm atmCC}_i+N^\text{\rm DSNB}_i +N^\text{\rm atmNC}_i}} \right)^2 \\
    + \left(\frac{\alpha}{\sigma_\alpha}\right)^2
  + \left(\frac{\beta}{\sigma_\beta}\right)^2
  + \left(\frac{\gamma}{\sigma_\gamma}\right)^2\ .
\end{multline}

\begin{figure}[!bt]
\centering
\includegraphics[width=0.45\textwidth]{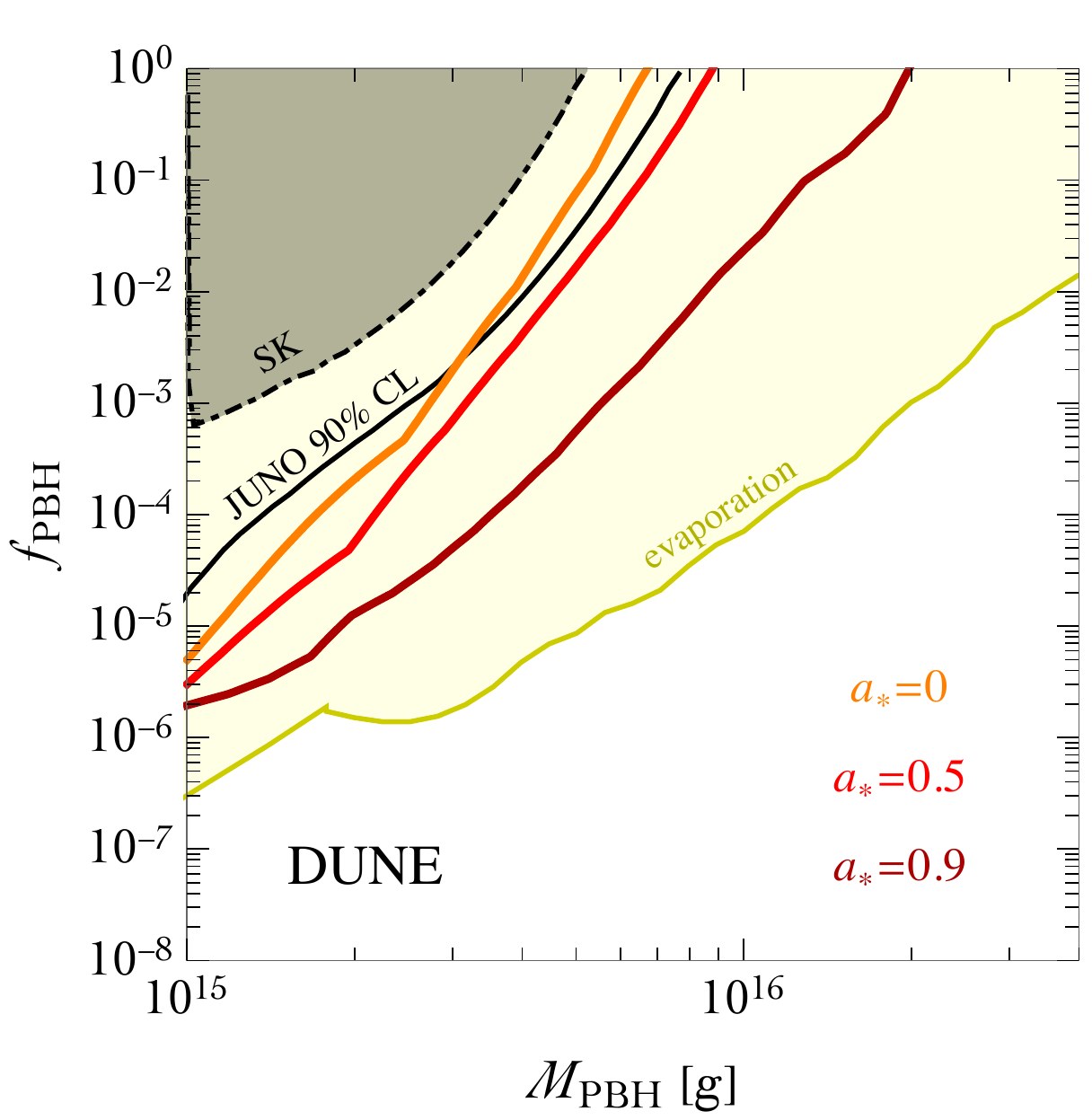}
\includegraphics[width=0.45\textwidth]{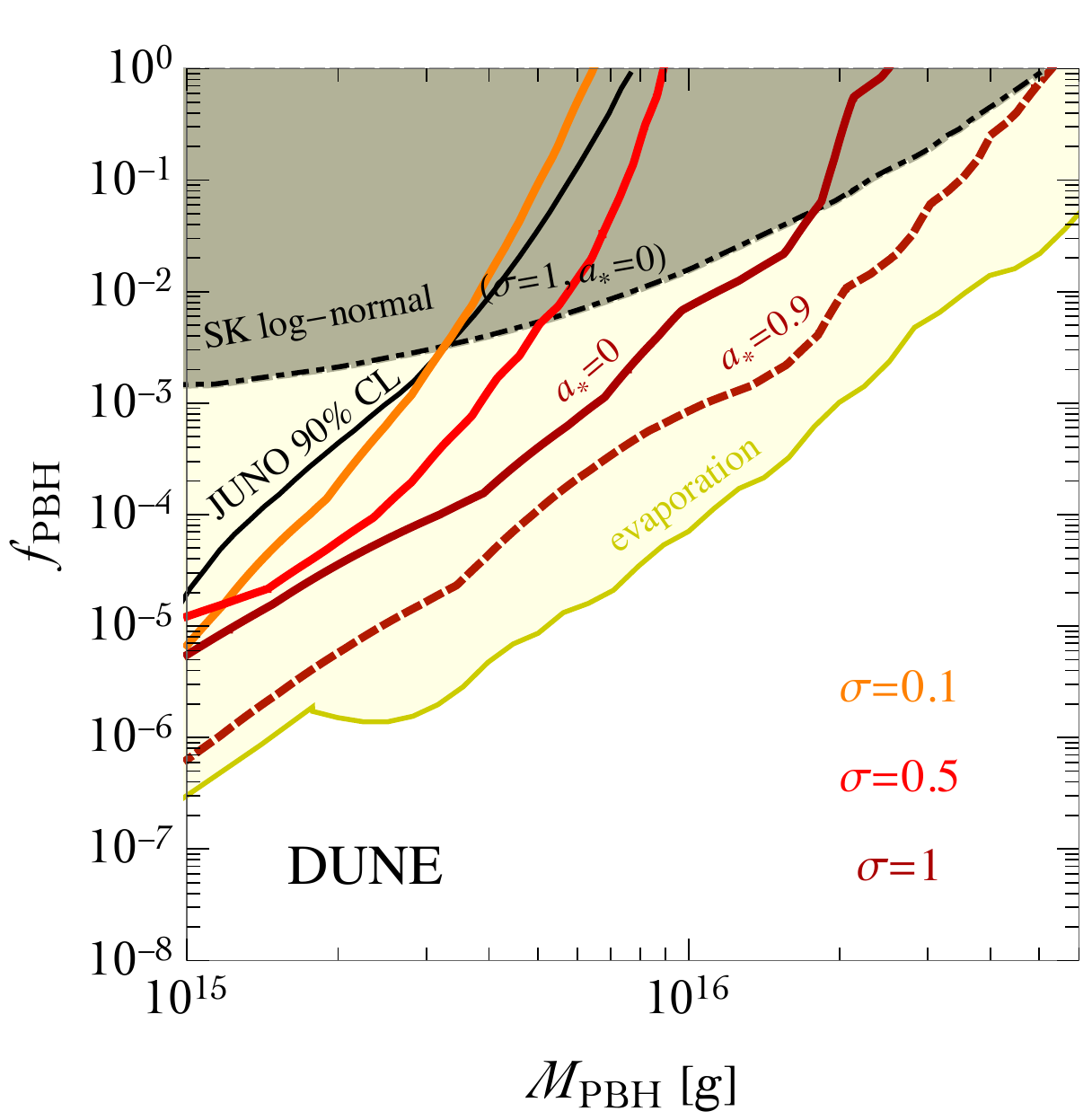}
\caption{Expected 95$\%$ C.L. sensitivities on the fraction of DM in the form of PBHs ($f_{\rm PBH}$) as a function of $M_{\rm PBH}$ at DUNE. The left panel assumes a monochromatic mass distribution of PBHs and three different spins. The right panel is for a log-normal PBH mass distribution with different widths. In this case, PBHs are assumed to be non-rotating except for the dark-red dashed contour which assumes PBHs with $a_*=0.9$.  For comparison, current Super-Kamiokande~\cite{Dasgupta:2019cae} and evaporation \cite{Clark:2016nst,Coogan:2020tuf,PBHbounds, bradley_j_kavanagh_2019_3538999} bounds (assuming a monochromatic mass distribution of non-spinning PBHs, unless otherwise labeled), together with the 90\% C.L. expected sensitivity at JUNO~\cite{Wang:2020uvi} are also shown. }
\label{fig:DUNEsens}
\end{figure}

The expected sensitivities for the DUNE experiment to a neutrino flux from PBH evaporation are shown in Figure \ref{fig:DUNEsens}. In the left panel, the 95$\%$ C.L. sensitivity curves (1 dof) on the PBH abundance $f_{\rm PBH}$ as a function of $M_{\rm PBH}$ are presented, for a monochromatic mass distribution of PBHs with different spins (assuming a monochromatic distribution also for $a_*$): $a_*=0$ (orange contour), $a_*=0.5$ (red) and $a_*=0.9$ (dark red). For comparison, the current upper limit from Super-Kamiokande~\cite{Dasgupta:2019cae} (black dot-dashed line) and from a combination of data (from \cite{PBHbounds, bradley_j_kavanagh_2019_3538999}, mainly extragalactic $\gamma-$ray background~\cite{Carr:2009jm,Ballesteros:2019exr,Arbey:2019vqx,Carr:2020gox}, CMB and MeV extragalactic $\gamma$-ray background~\cite{Clark:2016nst} and soft $\gamma$ rays from COMPTEL~\cite{Coogan:2020tuf}) together with the 90\% C.L. expected sensitivity at JUNO~\cite{Wang:2020uvi} (black, solid) are also shown. Unless labeled otherwise, all bounds and sensitivities quoted were derived assuming a monochromatic mass distribution of non-spinning PBHs. In all cases, the expected bounds will considerably improve the existing constraints set by Super-Kamiokande. For $a_*$ = 0, at masses below $\sim 3 \times 10^{15}$ g, the expected sensitivity of DUNE is also considerably better than for JUNO. This is due to the fact that DUNE is expected to have much better statistics given the significantly larger cross section. As the mass of the PBH increases, the overall flux decreases and it is shifted to lower energies. DUNE's low energy limit for these searches is fixed to 16 MeV, since at lower energies the solar neutrino flux would be orders of magnitude larger. In contrast, JUNO could extend the search down to approximately 12 MeV. Consequently, DUNE sensitivity is slightly worse at larger PBH masses. For rotating PBHs the particle emission is enhanced. Therefore, bounds for spinning primordial black holes turn out to be stronger. 

The predicted sensitivity of DUNE assuming an extended PBH mass distribution is shown in the right panel of Fig. \ref{fig:DUNEsens}. We consider a log-normal distribution for the PBH comoving number density (cf. Eq.~\ref{eq: log-normal mass distribution}) with different widths: $\sigma=0.1$ (orange contour), $\sigma=0.5$ (red) and $\sigma=1$ (dark red). In all cases we assume PBHs are non-rotating, except for the dark-red dashed contour, corresponding to the expected sensitivity for an  ``extreme" scenario with $\sigma=1$ and $a_*=0.9$.
For a narrow log-normal distribution ($\sigma = 0.1$) the sensitivity does not differ significantly from the monochromatic one with mass $M_{\rm PBH}=M_{\rm c}$. For wider distributions, however, the projected bounds are relaxed at low PBH masses but, on the other hand, slightly heavier PBHs can also be probed. This is a consequence of the change in the shape of the neutrino flux, which becomes broader and with its characteristic peak smoothed (see Fig. \ref{fig:lognorm}). The most stringent bound is expected for a particular scenario with a broad mass distribution of spinning PBHs ($\sigma=1$ and $a_*=0.9$). Also in this case, the predicted sensitivity of DUNE improves the current bound of Super-Kamiokande (for a log-normal mass distribution, with $\sigma=1$ and $a_*=0$ to be compared with our dark red solid line), in almost all parameter space except for $M_{\rm PBH} \gtrsim 10^{16}$ g. Overall, the fraction of DM in the form of PBHs that can be probed at DUNE is complementary to other evaporation constraints with different messengers (namely photons).\\

\begin{figure}[!tb]
\centering
\includegraphics[width=0.45\textwidth]{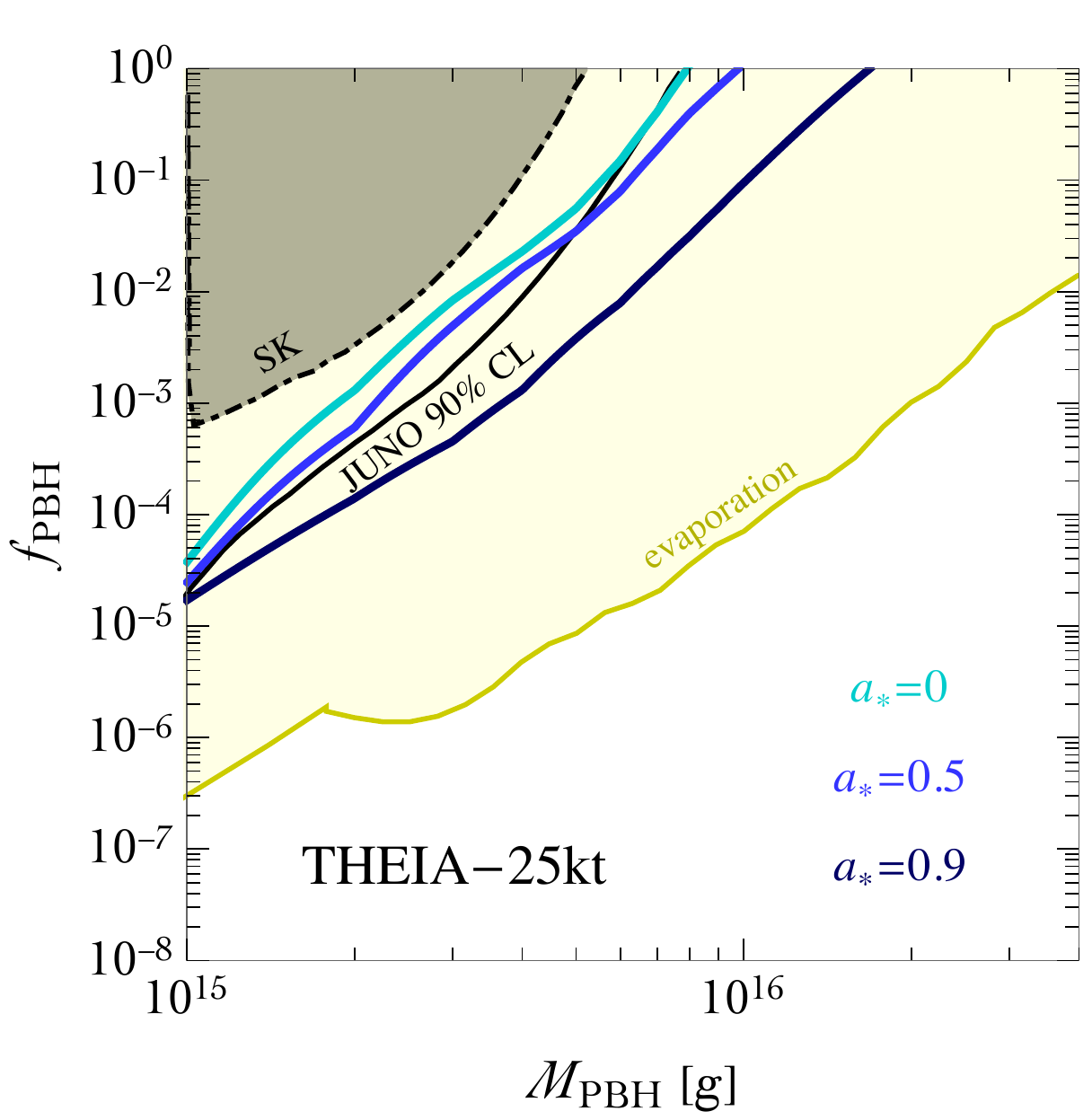}
\includegraphics[width=0.45\textwidth]{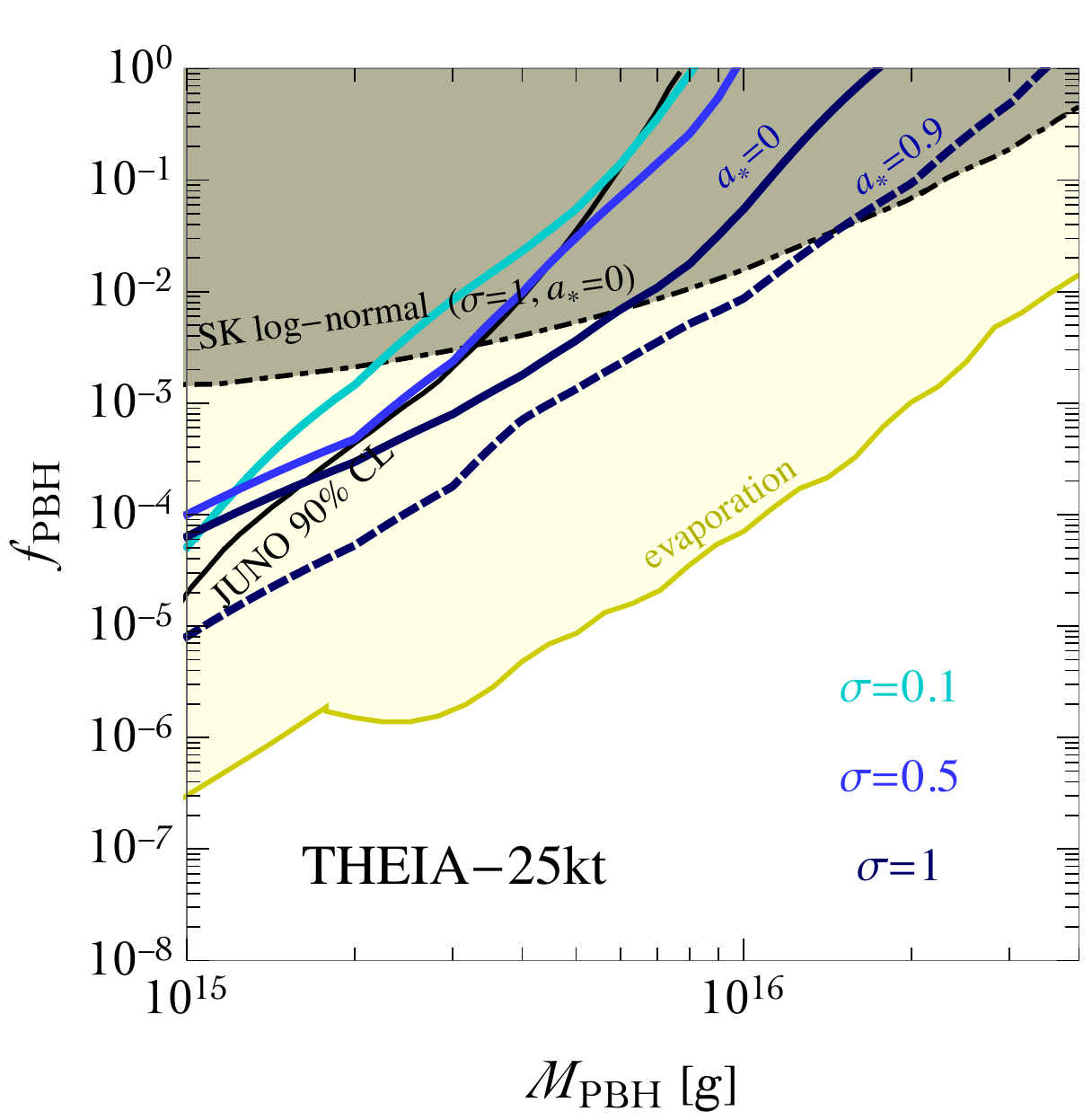}\\
\includegraphics[width=0.45\textwidth]{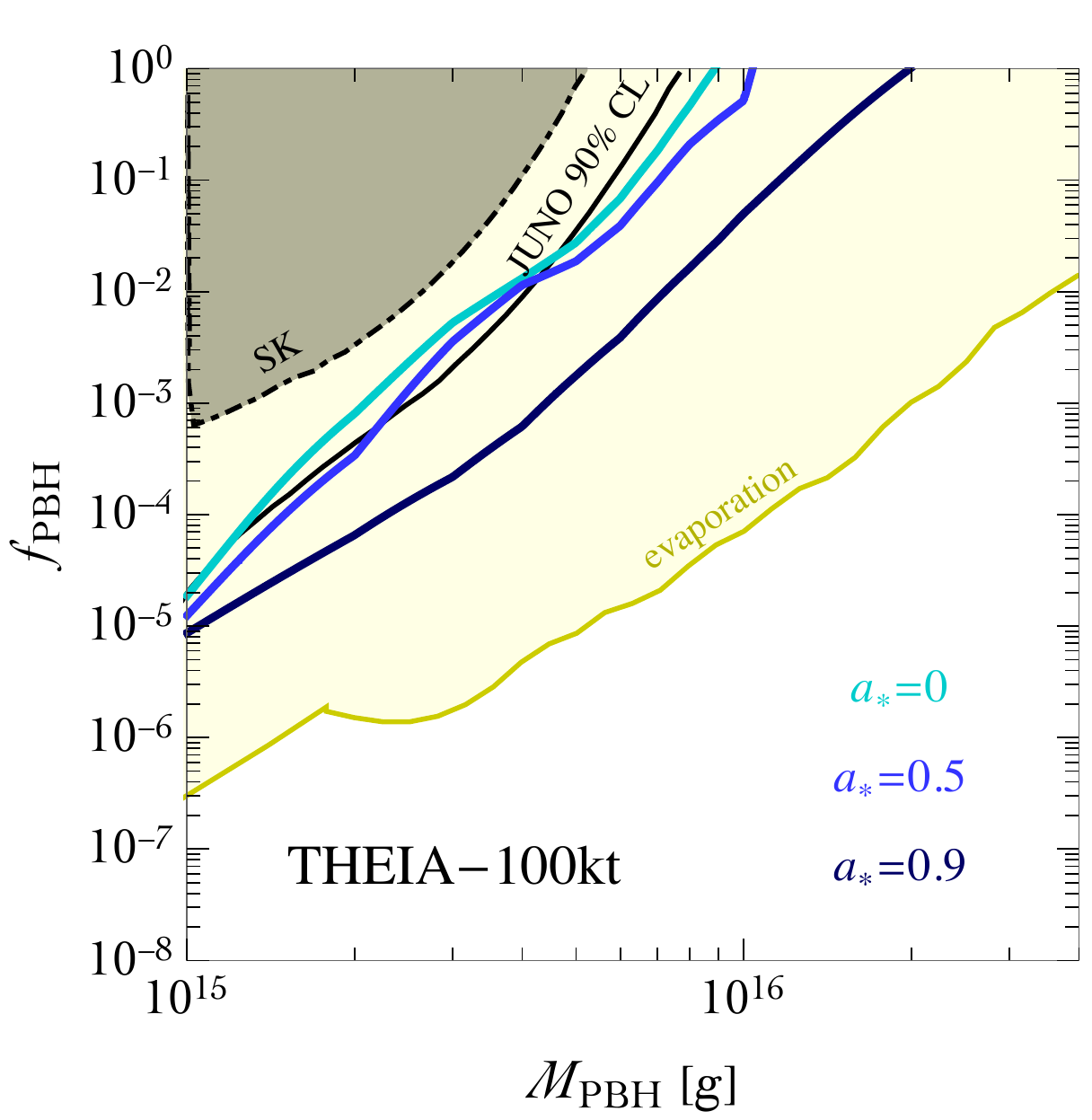}
\includegraphics[width=0.45\textwidth]{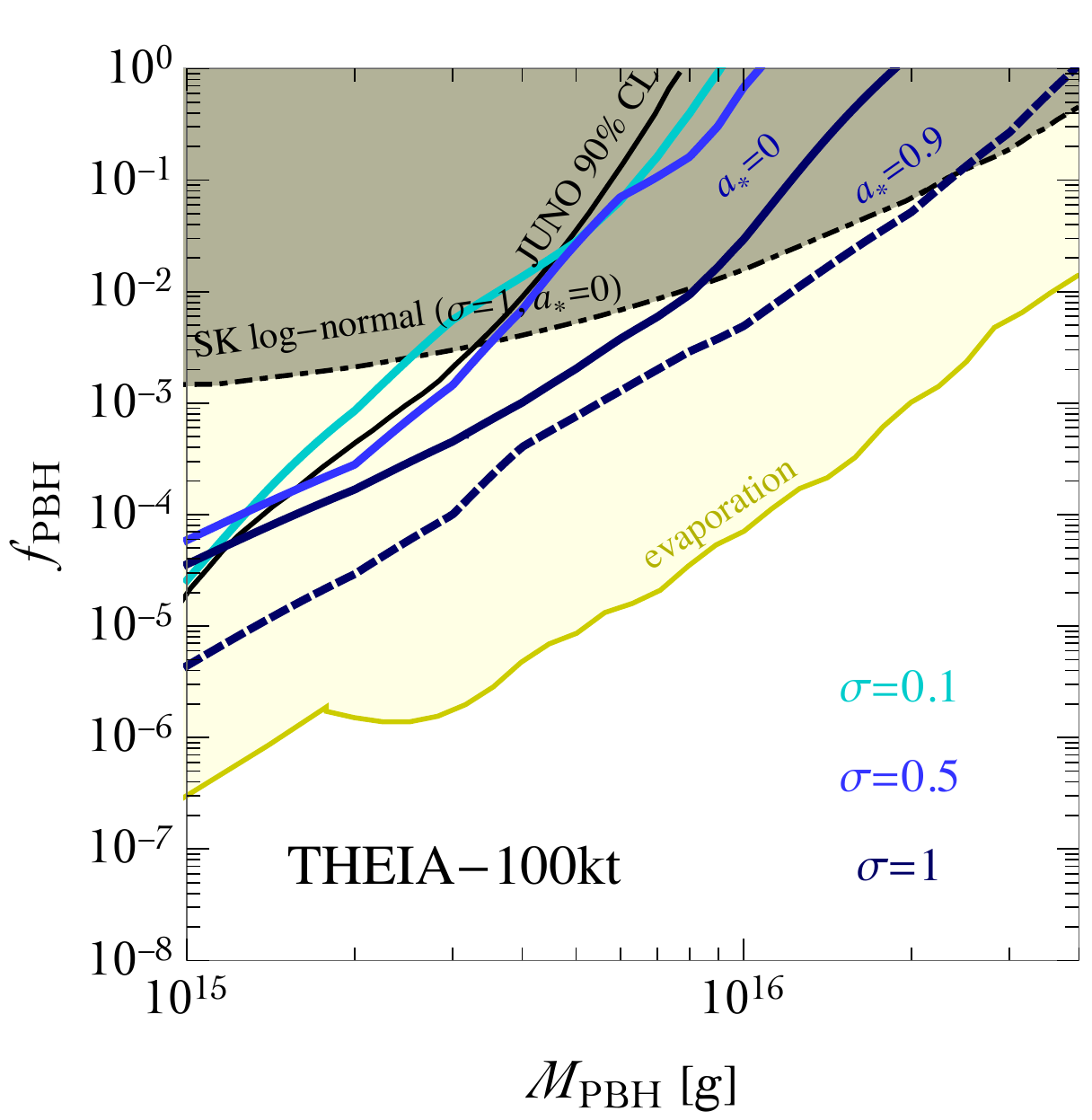}
\caption{Expected 95$\%$ C.L. sensitivities on the fraction of DM in the form of PBHs ($f_{\rm PBH}$) as a function of $M_{\rm PBH}$ at THEIA, assuming 20/80 kton of fiducial volume (upper/lower panels).  The left panels assume a monochromatic mass distribution of PBHs and three different spins. The right panels are for a log-normal PBH mass distribution with different widths. In this case, PBHs are assumed to be non-rotating except for the dark-blue dashed contour which assumes PBHs with $a_*=0.9$.  For comparison, current Super-Kamiokande~\cite{Dasgupta:2019cae} and evaporation \cite{Clark:2016nst,Coogan:2020tuf,PBHbounds, bradley_j_kavanagh_2019_3538999} bounds (assuming a monochromatic mass distribution of non-spinning PBHs, unless otherwise labeled) together with the 90\% C.L. expected sensitivity at JUNO~\cite{Wang:2020uvi} are also shown.}
\label{fig:THEIAsens}
\end{figure}

The future THEIA detector has the capability to provide complementary bounds to those of DUNE, since being a liquid scintillator it would be sensitive to the antineutrino component of the PBH flux. The expected sensitivity to rotating and non-rotating PBHs can be seen in the left panels of Figure \ref{fig:THEIAsens}. At low masses, bounds are comparable to those expected for JUNO~\cite{Wang:2020uvi}. Although the detection channel is the same, we could expect THEIA to have better statistics given the larger volume\footnote{Note that the neutrino flux close to the peak of the galactic component predicted in Ref.~\cite{Wang:2020uvi}  is about a factor 2 larger than our estimations in Fig.~\ref{fig:totaldFdE}.}. For higher masses, the antineutrino flux peaks at low energies around 10-20 MeV. Liquid scintillators are known to have a large background of atmospheric neutrinos mimicking the IBD signal. Nevertheless, WbLS detectors like THEIA can further minimise them, slightly enhancing the sensitivity to PBHs of masses larger than $\sim 8 \times 10^{15}$ g with respect to the sensitivity of a liquid scintillator. In the right panels of Figure \ref{fig:THEIAsens}, one can see the impact of considering a non-monochromatic distribution. Small departures from the monochromatic distribution (i.e. $\sigma = 0.1$) can not be distinguished. Nonetheless, for a broad mass distribution, the expected constraints for PBHs with masses close to $10^{15}$ g are weakened while, at the same time, the sensitivity to masses around $10^{16}$ g is improved. Finally, comparing the upper and lower panels in Fig.~\ref{fig:THEIAsens}, one can notice that the main difference between the two possible configurations of THEIA (25 kton and 100 kton), if in a similar location, is approximately a factor 2 in $f_{\rm PBH}$, due to the difference in statistics.

\section{Conclusions}
\label{sec:conclusions}
Primordial black holes are a viable dark matter candidate, whose masses can span over several orders of magnitude. Light PBHs with masses in the range $10^{15}-10^{17}$ g would evaporate and produce sizeable fluxes of MeV neutrinos via Hawking radiation. 
In this work, we have explored the possibility of detecting neutrinos from PBH evaporation with the future neutrino experiments DUNE and THEIA. 
We have investigated how the expected neutrino flux would change depending on the PBHs mass and spin, as well as the fraction of DM abundance they compose. In addition, we have considered  monochromatic as well as extended PBH mass distributions. 
Our analysis shows that both DUNE and THEIA could considerably improve the existing bounds from Super-Kamiokande on the abundance of PBHs with masses between $10^{15}-10^{16}$ g, obtained for a PBH monochromatic mass distribution, and even allow to probe heavier PBHs up to $\sim 5 \times 10^{16}$ g. 
If the PBHs follow a monochromatic mass distribution, DUNE and THEIA will be able to exclude non-rotating PBHs as the sole component of DM up to masses $7 \times 10^{15}$ g and  $9 \times 10^{15}$ g, respectively. The projected sensitivities are particularly good for masses around $10^{15}$ g, where abundances as low as $f_{\rm PBH}\sim$ few $\times 10^{-6}$ can be tested. Moreover, we find that if PBHs are assumed to have a non-zero spin, the derived sensitivities on their abundance are stronger. This is due to an enhancement in the neutrino flux and the appearance of characteristic features in the event spectrum when a rotating black hole is considered. In contrast, for an extended (log-normal) mass distribution, larger masses can be explored, although the sensitivity to light PBHs gets worse, as a consequence of the change in the flux shape.  

Eventually, our projections indicate that these next-generation neutrino facilities could set competitive bounds on the PBH parameter space, with respect to other neutrino constraints. It is interesting to see that they will also be complementary to existing limits 
from other cosmic messengers, depending on the underlying assumptions regarding the mass distribution and spins of the PBH population. 

In addition, DUNE and THEIA would complement each other since they are sensitive to neutrino and antineutrino fluxes from PBH evaporation, respectively.
Finally, let us stress that the neutrino signatures expected from PBH evaporation lay in the same energy range in which the diffuse supernova neutrino background is expected.
Consequently, the quest for PBH dark matter at future neutrino experiments like DUNE and THEIA could be easily included in their physics programs, leveraging existing supernova neutrino analyses.


\section*{Acknowledgments}
We are grateful to Sergio Palomares-Ruiz and Pablo Villanueva-Domingo for useful comments on this manuscript.
This work has been supported by the Spanish grants FPA2017-85216-P (AEI/FEDER, UE) and PROMETEO/2018/165 (Generalitat Valenciana).
VDR acknowledges financial support by the SEJI/2020/016 grant (project ``Les Fosques'') funded by
Generalitat Valenciana and by the Universitat de Val\`encia through the sub-programme “ATRACCI\'O DE TALENT 2019”. 
PMM is supported by the grant FPU18/04571.

\bibliographystyle{utphys}
\bibliography{bibliography}

\end{document}